# THE ABELL CLUSTER INERTIAL FRAME

Matthew Colless
Mount Stromlo and Siding Spring Observatories, The Australian National University, Weston Creek, ACT 2611, Australia

## ABSTRACT

This paper presents a re-analysis of Lauer & Postman's (1994) finding that the Abell cluster inertial frame (ACIF), defined by the 119 Abell clusters within 15,000 km s$^{-1}$, is moving at almost 700 km s$^{-1}$ with respect to the cosmic microwave background. Such a motion is inconsistent with most cosmological models at a confidence level of 95% or higher. We examine the use of the relation between the metric luminosity of brightest cluster galaxies and the slope of their luminosity profiles as an estimator of distances and peculiar velocities. We obtain an exact expression for a cluster's peculiar velocity in terms of the residual magnitude about this relation and compare this to the approximation used by Lauer & Postman. We critically examine the method used by Lauer & Postman to recover the Local Group motion from the scatter in this relation, and develop improved procedures including a maximum likelihood method that provides a direct estimate of the uncertainty in the derived motion. Simulations show this method yields an unbiased estimate for the Local Group motion with significantly smaller uncertainties than LP's original method. We re-analyse Lauer & Postman's data to obtain an improved estimate for the motion of the Local Group. We find that the Local Group is moving relative to the ACIF at $626 \pm 242$ km s$^{-1}$ towards $l$=216°, $b$=−28° (±20°). This implies that the ACIF is itself moving relative to the cosmic microwave background at $764 \pm 160$ km s$^{-1}$ towards $l$=341°, $b$=49° (±20°). This motion is consistent with that derived by LP but has a 10% larger amplitude and 20% smaller uncertainties, making it even harder to reconcile with cosmological models.

## 1. INTRODUCTION

Measuring the bulk motion of large volumes of the Universe has proved a powerful tool for testing cosmological models, since the amplitude of such motions is directly related to the power spectrum of density fluctuations (Peebles 1993). With the development of more precise distance estimators in recent years the size of the volumes over which peculiar velocities can be measured has steadily increased (Burstein 1990, Scaramella et al. 1994, Dekel 1994). Now Lauer & Postman (1994; hereafter LP) have measured the motion of the Local Group with respect to (w.r.t.) the inertial frame defined by the mean motion of the 119 Abell clusters closer than $cz$=15,000 km s$^{-1}$. They find that the Local Group is moving at almost 600 km s$^{-1}$ w.r.t. this Abell cluster inertial frame (ACIF) and in a direction that is 75° away from the direction in which the Local Group is moving w.r.t. the cosmic microwave background (CMB) frame (assuming that the CMB dipole is indeed due to relative motion). This implies that the ACIF is itself moving w.r.t. the CMB frame at almost 700 km s$^{-1}$.

That so large a volume of the Universe should be moving at such a high velocity w.r.t. the cosmic rest frame defined by the CMB is an intuitively surprising result. It implies that the CMB dipole must be generated by mass fluctuations at distances beyond those of the clusters in the ACIF, i.e. greater than 150 h$^{-1}$ Mpc. As recent calculations and simulations show (Feldman & Watkins 1994, Strauss et al. 1994), the power spectra of all the usual cosmological models predict that such a large motion for the ACIF has a probability of less than 5%.

Testing the correctness of this very important result is clearly essential. Several programs to do this are in progress: Lauer and Postman, with M.A.Strauss and J.Huchra, have begun work extending the cluster sample out to 24,000 km s$^{-1}$, increasing the sample size as well as the scale on which the bulk motion can be measured; the EFAR project (Colless et al. 1993) will obtain bulk motions in two superclusters using the $D_n$–$\sigma$ distance estimator for ellipticals; similar studies are also being carried out using the Tully-Fisher relation for spirals. However it is also necessary to confirm the methodology used to derive the bulk motions from such observations. This paper examines some of the methodological issues using LP's original dataset.

LP use the $M$–$\alpha$ relation between the metric luminosity and the luminosity profile slope of brightest cluster galaxies (BCGs) as the estimator of distances and peculiar velocities for clusters in the ACIF. We show that LP's expression for a BCG's peculiar velocity in terms of its residual with respect to the $M$–$\alpha$ relation is an approximation, and derive the exact relation. We also examine the physical basis for LP's procedure of minimising the rms BCG peculiar velocity as a means of inferring the Local Group motion in the case where the magnitude residuals are dominated by the intrinsic errors. We conclude that in these circumstances a straightforward minimisation of the magnitude residuals themselves is to be preferred. Finally, we re-analyse LP's original dataset using three different (though closely-related) methods to recover the Local Group motion: a revised version of their scheme employing minimisation of the rms BCG peculiar velocity, direct minimisation of the magnitude residuals, and a maximum likelihood scheme that allows estimation of confidence intervals without requiring Monte Carlo simulations of the data. We obtain a Local Group motion that is consistent with that derived by LP, but our improved methodology yields smaller uncertainties that increase the significance of this result.

## 2. THE DISTANCE ESTIMATOR

In this section we re-derive the distance (and peculiar velocity) estimator based on the use of the relation between metric magnitude and structure parameter $\alpha$ for brightest cluster galaxies (BCGs).

We consider our distance and motion w.r.t. a cluster in the ACIF. The ACIF is defined by the mean motion of the clusters it comprises. The redshift $z$ that we measure to one specific cluster may be decomposed as the product of three redshift components:

$$(1+z) = (1+z_H)(1+z_c)(1+z_p). \tag{1}$$

Here $z_H$ is the 'Hubble redshift' that we would measure if neither we nor the cluster had any peculiar velocity w.r.t. the ACIF (i.e. $z_H$ is the redshift corresponding to the cluster's true Hubble distance), $z_c$ corresponds to the peculiar velocity of the cluster w.r.t. the ACIF, and $z_p$ corresponds to the peculiar motion of the Local Group w.r.t. the ACIF. Note that for a particular cluster with direction given by the unit vector $\hat{\boldsymbol{r}}$, $z_p$ is related to the Local Group peculiar velocity $\boldsymbol{v}_p$ by

$$cz_p = -\boldsymbol{v}_p \cdot \hat{\boldsymbol{r}}, \tag{2}$$

where the minus sign results from the fact that if the Local Group motion is in the direction of the cluster then the cluster's redshift w.r.t. the Local Group is reduced.

We want to use the absolute luminosity of the BCG inside some specified metric radius, $L(r_0)$, as a standard candle distance estimator for the cluster. The angle, $\phi_0$, that $r_0$ subtends for the observer is given by the angular diameter distance of the cluster,

$$d_A \equiv r_0/\phi_0, \tag{3}$$

while the apparent luminosity within this angle, $l(\phi_0)$, is given by the luminosity distance,

$$d_L \equiv \left(\frac{L(r_0)}{4\pi l(\phi_0)}\right)^{1/2}. \tag{4}$$

These distances are related to the Hubble redshift of the cluster by

$$d_A(z_H) = (1+z_H)^{-2} d_L(z_H) \tag{5}$$

and

$$d_L(z_H) = \frac{cz_H}{H_0}\left[\frac{1+z_H+(1+2q_0z_H)^{1/2}}{1+q_0z_H+(1+2q_0z_H)^{1/2}}\right]. \tag{6}$$

We wish to measure $L(r_0)$, which means measuring $l(\phi_0)$. However we require some estimate for $z_H$ in order to determine $\phi_0$. Since our peculiar velocities are defined w.r.t. the ACIF, which is in turn defined by the mean motion of the clusters it comprises, the mean value of $z_c$ is zero and we may reasonably assume that $z_c \ll z_H$. Thus for a given value of $z_p$ we can use the approximation

$$(1+z_H) \approx (1+z)/(1+z_p) \equiv (1+z'). \tag{7}$$

Using $z'$ instead of $z_H$ corresponds to measuring the luminosity, $l(\phi)$, inside the angle

$$\phi = r_0/d_A(z') \equiv r/d_A(z_H). \tag{8}$$

Thus we measure the luminosity inside a metric radius $r$ rather than $r_0$.

If around $r_0$ the BCG's luminosity profile can be approximated by a power law of slope $\alpha$, then the ratio of the luminosity we actually measure to the one we seek is

$$\begin{aligned}
\frac{L(r)}{L(r_0)} &= \left(\frac{r}{r_0}\right)^{\alpha} \\
&= \left(\frac{d_A(z_H)}{d_A(z')}\right)^{\alpha} \\
&= \left(\frac{d_L(z_H)}{d_L(z')}\right)^{\alpha} \left(\frac{1+z'}{1+z_H}\right)^{2\alpha} \\
&= \left(\frac{d_L(z_H)}{d_L(z')}\right)^{\alpha} (1+z_c)^{2\alpha}
\end{aligned} \tag{9}$$

Now in practice we measure an apparent magnitude $m(\phi)$, which we have to convert to an absolute magnitude $M(r)$. Absolute and apparent magnitudes are related by

$$M(r) = m(\phi) - 5\log(d_L(z_H)) - K(z) - A - 25 \tag{10}$$

where $K(z)$ is the K-correction for a galaxy at measured redshift $z$, and $A$ is the absorption. Not knowing $z_H$, of course, we have to use $z'$, so we obtain

$$\begin{aligned}
M'(r) &= m(\phi) - 5\log(d_L(z')) - K(z) - A - 25 && (11)\\
&= M(r) - 5\log(d_L(z')/d_L(z_H)). && (12)
\end{aligned}$$

The difference between this observed absolute magnitude and the true metric magnitude is

$$\begin{aligned}
M'(r) - M(r_0) &= M(r) - M(r_0) - 5\log(d_L(z')/d_L(z_H)) \\
&= -2.5\log(L(r)/L(r_0)) - 5\log(d_L(z')/d_L(z_H)) \\
&= -2.5\alpha\log(d_L(z_H)/d_L(z')) - 5\alpha\log(1+z_c) - 5\log(d_L(z')/d_L(z_H)) \\
&= -2.5(2-\alpha)\log(d_L(z')/d_L(z_H)) - 5\alpha\log(1+z_c).
\end{aligned} \tag{13}$$

If we have assumed the correct Local Group motion then the above equation and the definition of $z'$ tell us that $M'(r) - M(r_0)$ is only non-zero (in the absence of errors) if $z_c$ is non-zero. Since by definition the mean value of $z_c$ is zero, we expect that $M'(r)$ will be very close to $M(r_0)$ in the mean.

In fact in order to take advantage of the known correlation between $M(r_0)$ and $\alpha$ and thus have a better standard candle, we use as our estimate of the true metric magnitude the mean relation $\overline{M'(r,\alpha)}$ evaluated at $\alpha(\phi)$, assuming that this is a good estimator of $M(r_0,\alpha)$. Thus the relation we use to estimate distances and peculiar velocities is

$$\Delta M \equiv M'(r) - \overline{M'(r,\alpha)} = -2.5(2-\alpha)\log(d_L(z')/d_L(z_H)) - 5\alpha\log(1+z_c). \qquad (14)$$

For an assumed Local Group motion $\boldsymbol{v}_p$ and observed values of $z$, $\Delta M$ and $\alpha$, we recover $z_H$ and $z_c$ by simultaneously solving this equation and equation 1.

This estimate of distance and peculiar velocity differs from that used by LP in two ways: (i) they omit the term $5\alpha\log(1+z_c)$ in equation 14, originating in the distinction between $d_A$ and $d_L$ (see equation 9); (ii) they use the approximation

$$d_L(z')/d_L(z_H) \approx cz'/cz_H \approx 1 + cz_c/cz_H \approx 1 + cz_c/cz \qquad (15)$$

in deriving their distance estimator. (It should be noted, however, that LP do use the correct cosmological expressions for the angular diameter distance and luminosity distance in computing metric apertures and luminosities.) Making these changes to the expression for $\Delta M$ gives

$$\Delta M \approx -2.5(2-\alpha)\log(1 + cz_c/cz). \qquad (16)$$

from which LP obtain their expression for the cluster's peculiar velocity:

$$cz_c = cz\left(\mathrm{dex}\left[\frac{-0.4\Delta M}{2-\alpha}\right] - 1\right). \qquad (17)$$

Figure 1 shows the difference, as a function of $\Delta M$, between the exact value of $cz_c$ (obtained by solving equations 1 and 14) and LP's approximate value (given by equation 17) for an illustrative case. We assume $H_0$=80 km s$^{-1}$ Mpc$^{-1}$ (though this does not enter the comparison) and $q_0$=0.5, and assume the cluster is at about the mean redshift of the ACIF sample, $z$=0.03, and the BCG has a typical structure parameter, $\alpha$=0.6. We plot the relations for $cz_p$=±600 km s$^{-1}$, roughly the amplitude of the Local Group motion w.r.t. the ACIF found by LP. The top panel of the figure shows the value of $cz_c$ inferred from a given value of $\Delta M$ using the exact expression ($cz_p$=+600 km s$^{-1}$ solid line; $z_p$=−600 km s$^{-1}$ dashed line) and LP's approximation (thick solid line; the two values of $cz_p$ give identical relations in LP's approximation). As the middle panel shows, the difference $\Delta cz_c$=$cz_c$(LP)−$cz_c$(true) can be large (several hundred km s$^{-1}$) for values of $\Delta M$ that are well within the range found by LP (solid line shows the difference for $cz_p$=+600 km s$^{-1}$, dashed line for $cz_p$=−600 km s$^{-1}$). The bottom panel shows the fractional difference, which ranges up to 50% for observed values of $\Delta M$.

Using LP's approximation to infer peculiar velocities *for individual clusters* can thus result in significant errors which are a function of the Local Group motion. However LP do not use their distance indicator for individual objects, but instead use it to derive the Local Group motion w.r.t. the ensemble of clusters making up the ACIF. By averaging over this all-sky sample the errors resulting from the use of the approximation cancel in first order, and the residual bias can be corrected for using Monte Carlo simulations. In the following section we examine the method used by LP to derive the Local Group motion and consider alternative approaches.

## 3. ESTIMATING THE LOCAL GROUP PECULIAR VELOCITY

The method LP use to derive the Local Group peculiar motion is to note that the rms peculiar velocity of the clusters in the ACIF will be minimised if it is computed using the true value of the Local Group motion. If an *incorrect* Local Group motion is assumed in computing the cluster peculiar velocities, this will add some spurious peculiar velocity $\Delta cz_p$ to $cz_c$, so that equation 17 becomes

$$\begin{aligned} cz_c &\approx cz\left(\mathrm{dex}\left[\frac{-0.4\Delta M}{2-\alpha}\right]-1\right)-\Delta cz_p \\ &\approx cz\left(\mathrm{dex}\left[\frac{-0.4\Delta M}{2-\alpha}\right]-1\right)+\Delta \boldsymbol{v}_p\cdot\hat{\boldsymbol{r}}. \end{aligned} \tag{18}$$

For the values of $\Delta M$ and $\alpha$ derived from this incorrect value of $\boldsymbol{v}_p$, LP find the value of $\Delta\boldsymbol{v}_p$ that minimises the rms cluster peculiar velocity $v_c$, given by

$$v_c = \left[\frac{\sum_i w_i (cz_{ci})^2}{\sum_i w_i}\right]^{1/2} \tag{19}$$

where the index $i$ runs over all the ACIF clusters and the $w_i$ are appropriate weights. They then update their estimate of $\boldsymbol{v}_p$ with this $\Delta\boldsymbol{v}_p$ and recompute $\Delta M$ and $\alpha$ for each object accordingly. The whole process is then repeated until $\Delta\boldsymbol{v}_p$ becomes negligibly small and the true value of $\boldsymbol{v}_p$ has been recovered. The weighting scheme preferred by LP has

$$w_i = (2-\alpha_i)^2/z_i^2, \tag{20}$$

which is intended to minimise the errors in $\boldsymbol{v}_p$ by allowing for the fact that more compact galaxies (with $\alpha \to 0$) and more nearby galaxies (with $z_i \to 0$) are more sensitive to $\boldsymbol{v}_p$.

The physical basis for this approach to recovering $\boldsymbol{v}_p$ rests upon the assumption that the residuals $\Delta M$ about the mean $M$–$\alpha$ relation are attributable to the peculiar velocities of the clusters w.r.t. the ACIF. However the observed rms residual of $\sigma_M$=0.24 mag, if interpreted in this way, would imply an implausible rms cluster peculiar velocity of about 1300 km s$^{-1}$. LP attempt to separate the intrinsic scatter in the $M$–$\alpha$ relation from that due to the cluster peculiar velocities by using the fact that the latter depends on redshift, and come to the unsurprising conclusion that the intrinsic scatter must greatly dominate the scatter due to cluster peculiar velocities.

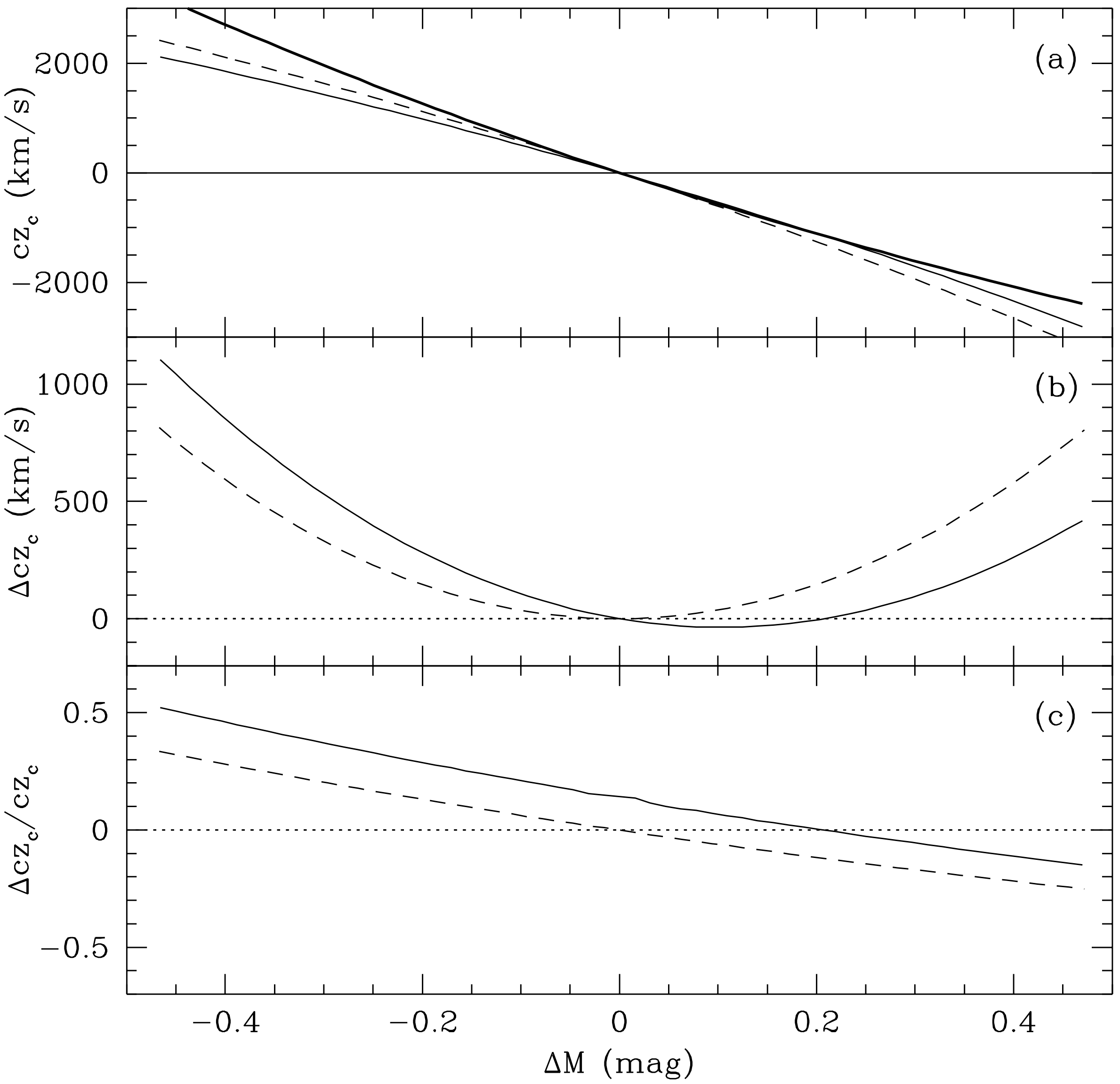

Fig. 1.— Comparison of the exact relation between $cz_c$ and $\Delta M$ with that obtained from LP's approximation. The comparison assumes the cluster is at $z$=0.03 and the BCG has $\alpha$=0.6 and shows the relations for $cz_p$=+600 km s$^{-1}$ (solid lines) and −600 km s$^{-1}$ (dashed lines): (a) $cz_c$ as a function of $\Delta M$ using the exact expression (thin lines) and LP's approximation (thick lines; the two values of $cz_p$ give identical relations in LP's approximation); (b) the difference $\Delta cz_c$=$cz_c$(LP)−$cz_c$(true); (c) the fractional difference $\Delta cz_c/cz_c$.

That being so, there is no physical reason for preferring to minimise $v_c$ rather than $\sigma_M$. As Figure 1 shows, $\Delta M$ and $cz_c$ are monotonically (and almost linearly) related. Hence minimising $\sigma_M$ over $\boldsymbol{v}_p$ should be very similar to minimising $v_c$ over $\boldsymbol{v}_p$, the only difference being that the relative weighting of clusters with different values of $\Delta M$ will differ to the extent that $cz_c$ is *not* linearly related to $\Delta M$. It is worth noting that, for small $\Delta M$ (i.e. small $cz_c$), LP's expression for $cz_c$ (equation 17) reduces to

$$cz_c = -0.4 \ln 10 \left( \frac{cz\Delta M}{2-\alpha} \right) . \tag{21}$$

Applying the weighting of equation 19 in computing the rms cluster peculiar velocity gives

$$v_c = 0.4c \ln 10 \left[ \frac{\sum_i \Delta M_i^2}{\sum_i (2-\alpha_i)^2 / z_i^2} \right]^{1/2} . \tag{22}$$

Since the observed redshifts $z_i$ are fixed and $\alpha_i$ changes very little with $\boldsymbol{v}_p$, minimising the rms cluster peculiar velocity w.r.t. $\boldsymbol{v}_p$ using these weights is, for small $\Delta M$, almost identical to minimising the unweighted rms magnitude residual.

An alternative to minimising $v_c$ or $\sigma_M$ is to use a maximum likelihood approach based on the observation that the residuals about the mean $M$–$\alpha$ relation have a Gaussian distribution (Postman & Lauer 1995). The probability of observing a given BCG's magnitude residual given an assumed $\boldsymbol{v}_p$ is $P_i = N(\Delta M : \delta M, \sigma_M)$ where $\Delta M$ is the observed residual about the mean *M-alpha* relation, $\delta M$ is the predicted residual given by the RHS of equation 14, $\sigma_M$ is the minimised scatter in the $M$–$\alpha$ relation and $N(x : \mu, \sigma)$ is the value at $x$ of a Gaussian of mean $\mu$ and dispersion $\sigma$. Computing $\delta M$ requires $z_c$, but we can make the simplifying approximation $\delta M$=0 (corresponding to $z_c$=0) because the intrinsic scatter completely dominates the cluster peculiar velocities, so that $\delta M \ll \sigma_M$ and hence $P_i \approx N(\Delta M : 0, \sigma_M)$.

Given this probability for each galaxy in the sample we can recover the Local Group motion by maximising the log-likelihood for the entire sample, $\ln \mathcal{L} = \sum_i \ln(P_i)$, as a function of the assumed value of $\boldsymbol{v}_p$. The particular advantage of this method is that it allows us to estimate uncertainties using the property of the likelihood that $2\ln(\mathcal{L}_{max}/\mathcal{L})$ is distributed approximately as $\chi^2_\nu$, with the number of degrees of freedom $\nu$ corresponding to the number of free parameters in the model. We can therefore use constant-$\chi^2$ boundaries to obtain reliable confidence intervals (uncertainties) for the model parameters both individually and jointly without recourse to Monte Carlo simulations (see, e.g., Press et al. 1986, pp533-7).

We can thus estimate the Local Group motion by minimising over $\boldsymbol{v}_p$ any one of four different (though intimately connected) quantities: (a) the weighted rms cluster peculiar velocity, $v_c(LP)$, using LP's approximation; (b) the weighted rms cluster peculiar velocity, $v_c$, using the exact expression; (c) the unweighted rms magnitude residual in the $M$–$\alpha$ relation, $\sigma_M$; and (d) the negative log-likelihood of the magnitude residuals, $-\ln\mathcal{L}$. The details of the procedure are as follows:

1. Adopt a Local Group peculiar velocity $\boldsymbol{v}_p$. For each cluster $i$ in the ACIF compute $z_{pi} = -\hat{\boldsymbol{r}}_i \cdot \boldsymbol{v}_p/c$ where $\hat{\boldsymbol{r}}_i$ is the unit vector in the direction of the cluster. Then compute $z_i' = (1+z_i)/(1+z_{pi}) - 1$.

2. Estimate the approximate angle subtended by the metric radius, $\phi_i = r_0/d_A(z_i')$, and measure the apparent magnitude $m_i(\phi)$ and $\alpha_i(\phi) = d\log L/d\log\phi\,|_{\phi_i}$ from the observed luminosity profile of the BCG.

3. Compute $M_i'(r)$ from equation 11. Fit the $M'(r)$ versus $\alpha$ relation with a quadratic and obtain the residuals about this mean relation, $\Delta M_i = M_i'(r) - \overline{M'(r,\alpha_i)}$.

4. For method (a): Compute the notional rms peculiar velocity of the ACIF clusters using LP's approximation and weights (equations 17, 19 and 20).

5. For method (b): Solve equations 1 and 14 simultaneously to derive $z_{Hi}$ and $z_{ci}$, and compute the notional rms peculiar velocity of the ACIF clusters from equations 19 and 20.

6. For method (c): Compute the rms residual $\sigma_M$ about the mean $M$–$\alpha$ relation.

7. For method (d): Compute the log-likelihood $\ln\mathcal{L} = \sum_i \ln(N(\Delta M\!:\!0,\overline{\sigma_M}))$, where $\overline{\sigma_M}$ is the minimum rms magnitude residual.

8. Repeat these steps, attempting to minimise $\sigma_M$, $v_c$ or $-\ln\mathcal{L}$ as a function of $\boldsymbol{v}_p$. The value of $\boldsymbol{v}_p$ that gives the minimum is the estimate of the Local Group peculiar velocity.

One further difference between this procedure and that used by LP is that we update the $\alpha_i$ and $\Delta M_i$ at every step in our minimisation rather than using a two-stage process of minimising $v_c$ over $\boldsymbol{v}_p$ for a *fixed* set of $\alpha_i$'s and $\Delta M_i$'s and then updating these quantities for the new estimate of $\boldsymbol{v}_p$ and repeating the process until it converges. Using the one-step procedure is conceptually neater, though it probably makes little difference to the final result.

We now apply these various methods to LP's data for the BCGs and clusters in the ACIF in order to see what differences result in the inferred Local Group motion.

## 4. RE-ANALYSING THE ACIF

In re-analysing the ACIF with the procedures described above, we use the original data on the Abell cluster sample and the BCGs given in Lauer & Postman (1994; LP) and Postman & Lauer (1995). Since some processing of the observed luminosity profiles is required to derive the values of $\alpha$, $M$ and $\Delta M$ for a given $\boldsymbol{v}_p$, the first step is to make a consistency check between our values and those obtained by LP.

Table 3 of LP lists the $\alpha$, $M$ and $\Delta M$ of each BCG for three assumed Local Group motions: (i) case C, in which the ACIF is assumed to be at rest w.r.t. the CMB, so that the Local Group velocity w.r.t. the ACIF is the same as the Local Group velocity w.r.t. the CMB, i.e. $\boldsymbol{v}_p=\boldsymbol{C}$; (ii) case L, in which the Local Group and ACIF are assumed to be at rest w.r.t. each other, i.e. $\boldsymbol{v}_p=0$; and (iii) case F, in which the Local Group has LP's derived velocity $\boldsymbol{L}$ w.r.t. the ACIF, i.e. $\boldsymbol{v}_p=\boldsymbol{L}$. (Note that case L does *not* correspond to $\boldsymbol{v}_p=\boldsymbol{L}$ and that case F does *not* correspond to $\boldsymbol{v}_p=\boldsymbol{F}$, the motion of the ACIF w.r.t. the CMB.) There is a minor ambiguity in recovering LP's results due to the fact that there are 4 clusters (407, 419, 1177 and 3574) for which the cluster redshift (computed using the value in Table 1 of Postman & Lauer (1995) transformed to the Local Group frame as specified in LP) is not identical to the value listed in Table 1 of LP. These differences are small (the largest is less than 100 km s$^{-1}$) and we have simply adopted the values from Postman & Lauer (1995) since they give somewhat better agreement with the $\alpha$, $M$ and $\Delta M$ values listed in Table 3 of LP.

Figure 2 shows the differences between our values of $\alpha$, $M$ and $\Delta M$ and LP's values for their three tabulated cases of $\boldsymbol{v}_p$. (Note that the values given by LP for case F are for their estimated Local Group velocity *without* the small correction for the bias introduced by the sample geometry which they apply to obtain their final $\boldsymbol{L}$.) Since we are attempting to reproduce LP's procedure for deriving these quantities, this is purely a test of whether we are successful in recovering their numbers, and is not a check on whether the assumed Local Group motion is correct. The differences are plotted as a function of $\cos\theta$, the cosine of the angle between the direction of the cluster and the direction of the assumed Local Group motion. For all three cases any systematic differences are negligible and the rms differences between our values and LP's are approximately 0.003 in $\alpha$, 0.001 mag in $M$ and 0.004 mag in $\Delta M$, corresponding to an rms difference in $cz_c$ of $\sim$25 km s$^{-1}$.

This comparison reassures us that we are recovering $\alpha$, $M$ and $\Delta M$ in a manner consistent with LP, and we can proceed to recover $\boldsymbol{v}_p$ using the methods described in the previous section. Table 1 gives the value of $\boldsymbol{v}_p$ obtained using the four methods described above and starting from three initial estimates: $\boldsymbol{v}_{init}=0$, $\boldsymbol{v}_{init}=\boldsymbol{C}$ and $\boldsymbol{v}_{init}=-\boldsymbol{C}$. We used the `AMOEBA` routine from Numerical Recipes to do the minimisation, and demanded that the quantity being minimised converge to within a fractional tolerance of $10^{-4}$. The table lists both the initial velocity $\boldsymbol{v}_{init}$ and the recovered Local Group motion $\boldsymbol{v}_p$ in Cartesian Galactic coordinates. For this value of $\boldsymbol{v}_p$, the table gives the rms residual about the quadratic fit to the $M$–$\alpha$ relation ($\sigma_M$), the notional rms cluster peculiar velocity ($v_c$) and the log-likelihood of the magnitude residuals ($\ln\mathcal{L}$). Note that only one of these three quantities is actually optimised at a time; the others are simply the values at the optimum $\boldsymbol{v}_p$. The final column is the confidence level, $P(\mathcal{L})$, at which we can reject the solution based on it's likelihood ratio w.r.t. the maximum likelihood solution—this is a measure of the consistency of the solutions obtained from the different methods.

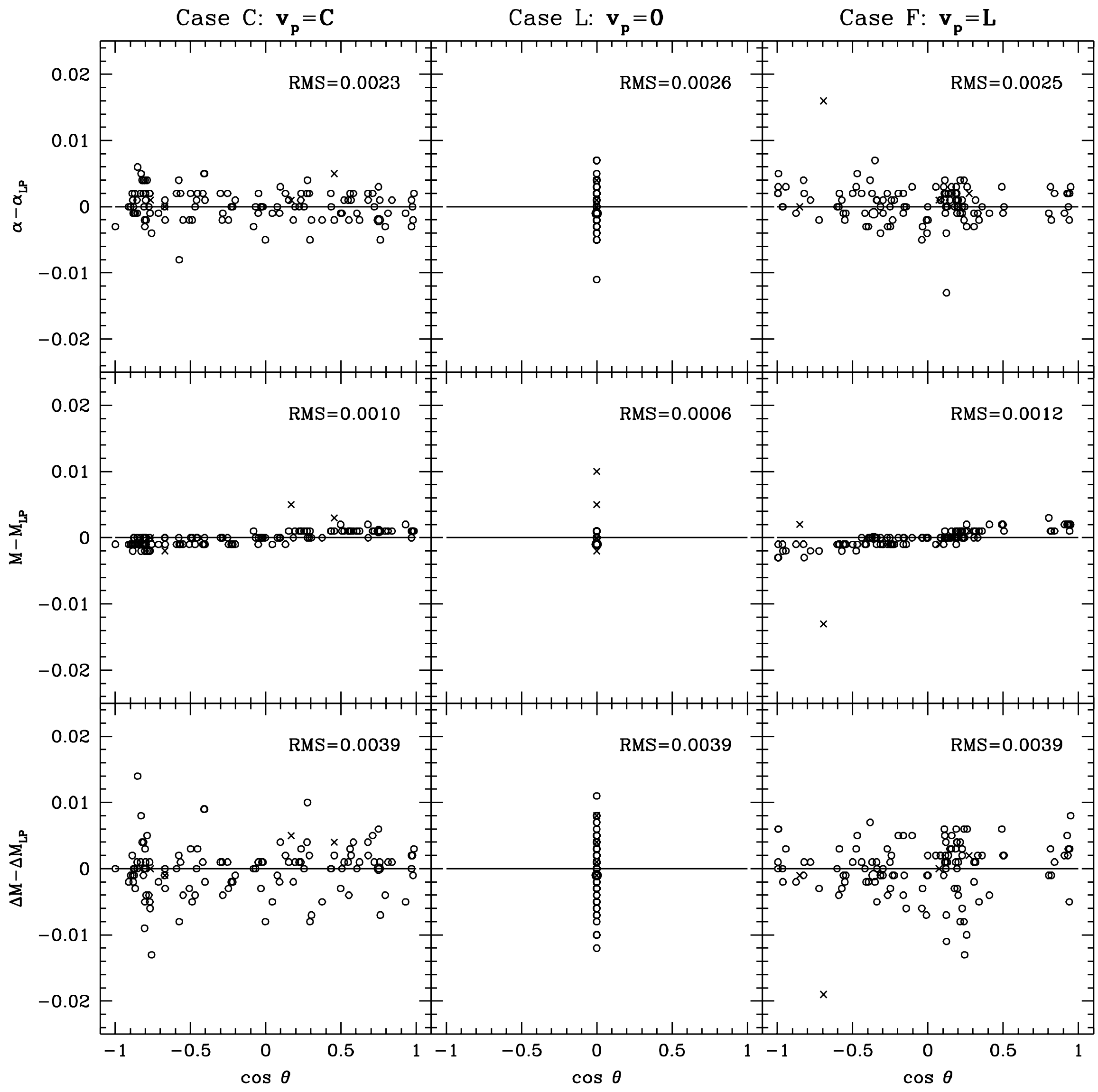

Fig. 2.— The differences between our derived values of $\alpha$, $M$ and $\Delta M$ and those tabulated in Table 3 of LP for three specific Local Group velocities: $\boldsymbol{v}_p$=$\boldsymbol{C}$, $\boldsymbol{v}_p$=0 and $\boldsymbol{v}_p$=$\boldsymbol{L}$. The differences are plotted as a function of $\cos\theta$, the cosine of the angle between the direction of the cluster and the direction of the assumed Local Group motion. The rms scatter is given excluding the four clusters (indicated by crosses) with slightly discrepant redshift values (see text).

TABLE 1

THE LOCAL GROUP PECULIAR VELOCITY

| $\boldsymbol{v}_{init}$ ( km s$^{-1}$) | $\boldsymbol{v}_p$ ( km s$^{-1}$) | $\sigma_M$ (mag) | $v_c$ ( km s$^{-1}$) | $\ln\mathcal{L}$ | $P(\mathcal{L})$ |
|---|---|---|---|---|---|
| (a) Minimising LP's $v_c$: | | | | | |
| ( 0, 0, 0) | (−535, −198, −298) | 0.243 | 1388 | -0.7268 | 20% |
| ( +7, −542, +302) | (−530, −211, −317) | 0.243 | 1388 | -0.6802 | 18% |
| ( −7, +542, −302) | (−543, −199, −315) | 0.243 | 1388 | -0.7896 | 23% |
| (b) Minimising $v_c$: | | | | | |
| ( 0, 0, 0) | (−567, −216, −308) | 0.243 | 1265 | -0.7641 | 22% |
| ( +7, −542, +302) | (−590, −229, −339) | 0.243 | 1266 | -0.8790 | 27% |
| ( −7, +542, −302) | (−615, −296, −296) | 0.243 | 1267 | -0.5820 | 13% |
| (c) Minimising $\sigma_M$: | | | | | |
| ( 0, 0, 0) | (−512, −368, −326) | 0.242 | 1269 | -0.2302 | <1% |
| ( +7, −542, +302) | (−490, −373, −342) | 0.242 | 1270 | -0.2364 | <1% |
| ( −7, +542, −302) | (−466, −364, −343) | 0.242 | 1270 | -0.2570 | <1% |
| (d) Maximising $\ln\mathcal{L}$: | | | | | |
| ( 0, 0, 0) | (−506, −367, −329) | 0.242 | 1269 | -0.2288 | — |
| ( +7, −542, +302) | (−508, −373, −328) | 0.242 | 1268 | -0.2290 | — |
| ( −7, +542, −302) | (−507, −371, −330) | 0.242 | 1269 | -0.2288 | — |

TABLE 2

COMPARISON WITH LAUER & POSTMAN

| | $V_x$ ( km s$^{-1}$) | $V_y$ ( km s$^{-1}$) | $V_z$ ( km s$^{-1}$) | $\lvert\boldsymbol{v}_p\rvert$ ( km s$^{-1}$) | $(l_p, b_p)$ (deg) | $\sigma_\theta$ (deg) | $\Delta\theta$ (deg) | $P(\mathcal{L})$ |
|---|---|---|---|---|---|---|---|---|
| $\boldsymbol{v}_p$ | $-506 \pm 188$ | $-367 \pm 213$ | $-329 \pm 163$ | $626 \pm 242$ | (216, −28) | 20 | — | — |
| $\boldsymbol{L}$ | $-470 \pm 250$ | $-399 \pm 273$ | $-333 \pm 198$ | $561 \pm 284$ | (220, −28) | 27 | 4 | 0.8% |
| $\boldsymbol{C}$ | $+7 \pm 8$ | $-542 \pm 15$ | $+302 \pm 15$ | $621 \pm 10$ | (271, +29) | 1 | 78 | >99.99% |
| 0 | 0 | 0 | 0 | 0 | — | — | — | 98.5% |

For each method the final value of $\boldsymbol{v}_p$ is similar for all values of $\boldsymbol{v}_{init}$; other values of $\boldsymbol{v}_{init}$, not shown in the table, were also tried and gave the same values of $\boldsymbol{v}_p$. As expected from the above discussion, the values of $P(\mathcal{L})$ show that all four methods give consistent results for $\boldsymbol{v}_p$ within the uncertainties, although there is a slight systematic difference between the methods minimising the cluster peculiar velocities and those minimising the magnitude residuals.

We estimate uncertainties from the maximum likelihood method. The 68% confidence interval for each component of $\boldsymbol{v}_p$=$(V_x, V_y, V_z)$ is the projection of the 3D likelihood contour $2(\ln\mathcal{L}_{max} - \ln\mathcal{L})$=1.0 onto that component's axis. We find that the Cartesian components of the best solution for $\boldsymbol{v}_p$ are

$$\begin{aligned} V_x &= -506 \pm 188 \text{ km s}^{-1} \\ V_y &= -367 \pm 213 \text{ km s}^{-1} \\ V_z &= -329 \pm 163 \text{ km s}^{-1} \end{aligned} \tag{23}$$

which corresponds to a Local Group motion $|\boldsymbol{v}_p| = 626 \pm 242$ km s$^{-1}$ towards $(l_p, b_p) = (216, -28)$ with a total angular error $\sigma_\theta = 20°$. Note that the radial amplitude of the Local Group motion has been corrected for error biasing according to $|\boldsymbol{v}_p|^2 = (V_x^2 + V_y^2 + V_z^2) - (\delta V_x^2 + \delta V_y^2 + \delta V_z^2) = 706^2 - 328^2$. Figure 3 shows how the rms cluster peculiar velocity $v_c$, the rms residual magnitude $\sigma_M$ and the log-likelihood $\ln(\mathcal{L})$ vary with the assumed *amplitude* $|\boldsymbol{v}_p|$ of the Local Group motion, while Figure 4 shows how these quantities vary with the assumed *direction* $(l_p, b_p)$ of the Local Group motion. The dipole signature of the Local Group motion is clearly seen in Figure 4; the extent to which the pole and anti-pole are not 180° opposed is consistent with the error estimates. The $M$–$\alpha$ relation for our best estimate of $\boldsymbol{v}_p$ is shown in Figure 5. The best-fit quadratic to this relation has an rms residual of 0.242 mag and coefficients $c_0$=−20.856, $c_1$=−4.529 and $c_2$=2.845. We confirm the finding of Postman & Lauer (1995) that there is no significant variation of the residuals with $\alpha$ or $z$, and that the distribution of residuals is well-represented by a Gaussian, as we assumed in constructing the likelihood.

Table 2 compares our maximum likelihood solution for $\boldsymbol{v}_p$ with the preferred solution of LP ($\boldsymbol{v}_p$=$\boldsymbol{L}$), the case in which the ACIF is at rest w.r.t. the CMB ($\boldsymbol{v}_p$=$\boldsymbol{C}$) and the case in which the Local Group is at rest w.r.t. the ACIF ($\boldsymbol{v}_p$=0). We list the components of these velocities (and their uncertainties) in both Cartesian and spherical Galactic coordinates. We also give the angle between our value of $\boldsymbol{v}_p$ and the alternate values, $\Delta\theta$, and the confidence level, $P(\mathcal{L})$, at which we can reject the alternate values based on their likelihood ratio w.r.t. our preferred solution.

The solution we obtain for $\boldsymbol{v}_p$ is entirely consistent with that of LP—the offset is just $\boldsymbol{L} - \boldsymbol{v}_p = (+36, -32, -4)$. Differences in the procedures for deriving $\boldsymbol{v}_p$ have not led to any significant differences in the final result. However it is worth pointing out that the agreement in $|\boldsymbol{v}_p|$ is slightly worse than might be expected from the very good agreement in $(V_x, V_y, V_z)$. This is entirely a consequence of the fact that we obtain roughly 20% smaller uncertainties for the components of $\boldsymbol{v}_p$ and so make a smaller correction for error biasing in computing $|\boldsymbol{v}_p|$. These smaller uncertainties mean that we rule out the alternatives $\boldsymbol{v}_p$=$\boldsymbol{C}$ and $\boldsymbol{v}_p$=0 with greater

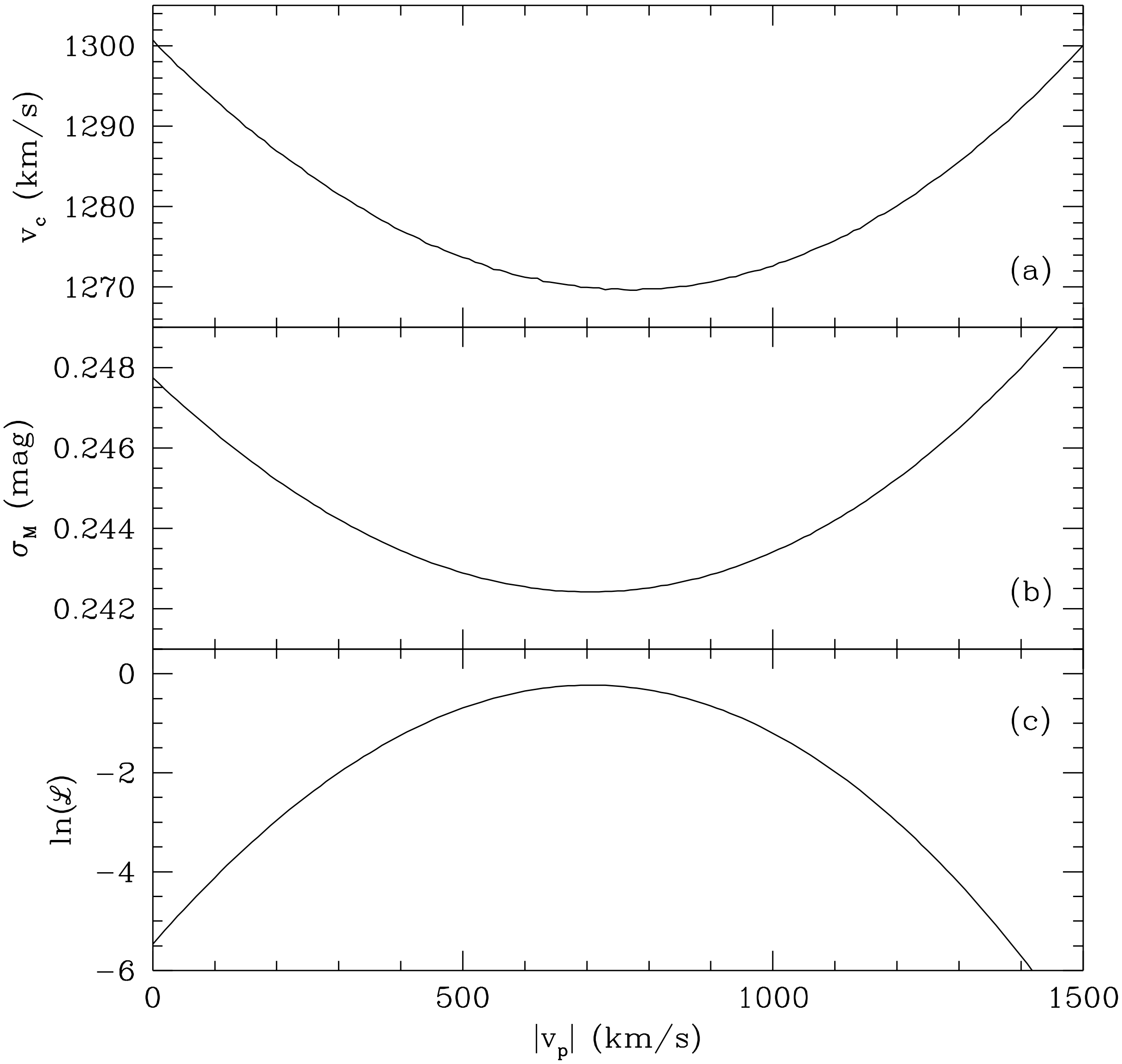

Fig. 3.— (a) The rms cluster peculiar velocity, (b) the rms residual magnitude, and (c) the log-likelihood, all as functions of $|\boldsymbol{v}_p|$ for an assumed direction of the Local Group motion of $(l_p, b_p)$=(216, −28).

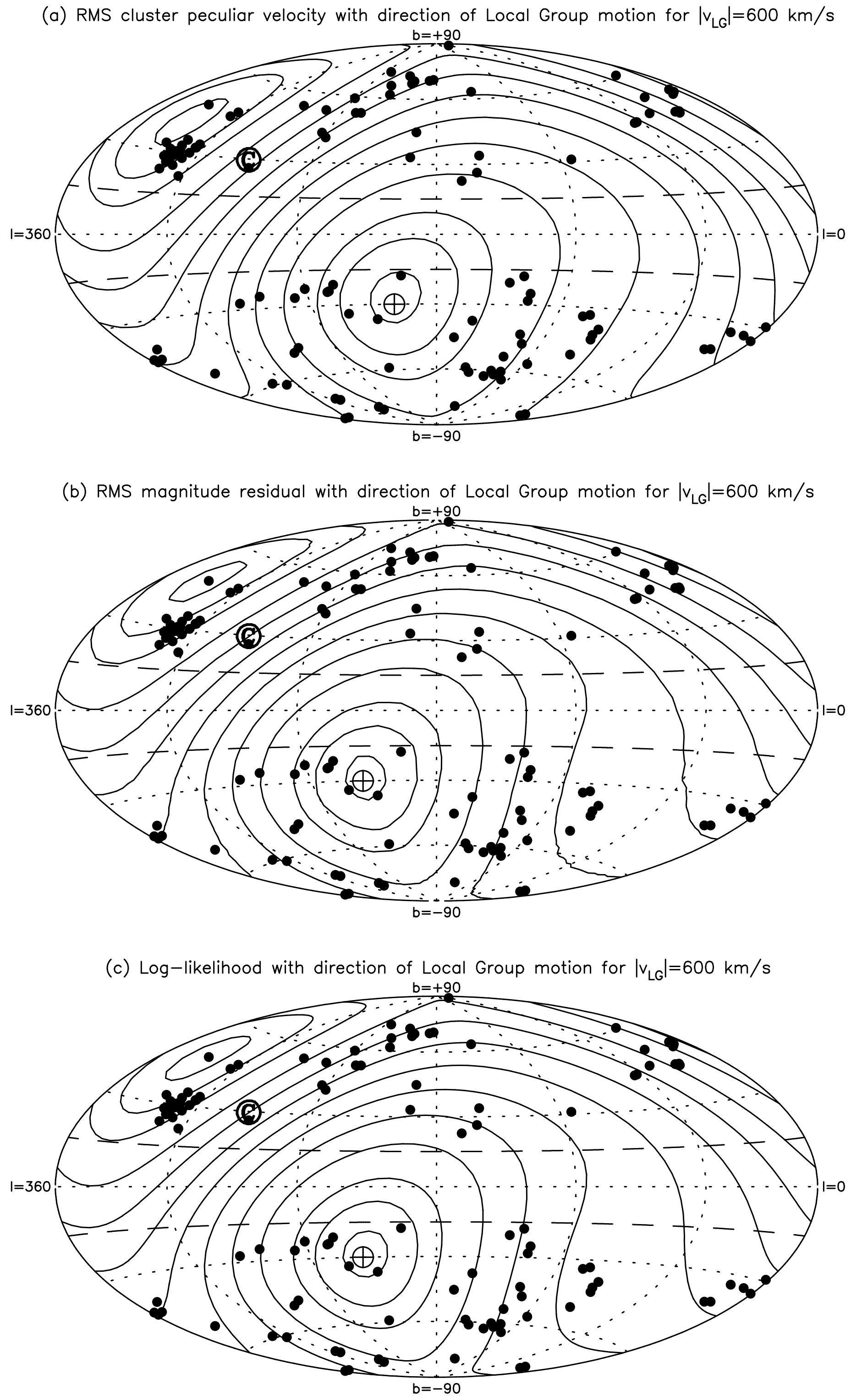

Fig. 4.— Contour plots of (a) the rms cluster peculiar velocity, (b) the rms residual magnitude, and (c) the log-likelihood, as functions of $(l_p, b_p)$ for an assumed amplitude of the Local Group motion of $|\boldsymbol{v}_p|$=600 km s$^{-1}$. The inferred direction of motion of the Local Group w.r.t. the ACIF is indicated as $\bigoplus$; the direction of motion of the Local Group w.r.t. the CMB is indicated as ©. The dots show the positions of the ACIF cluster sample. The zone of avoidance lies between the dashed lines at $b$=±15°.

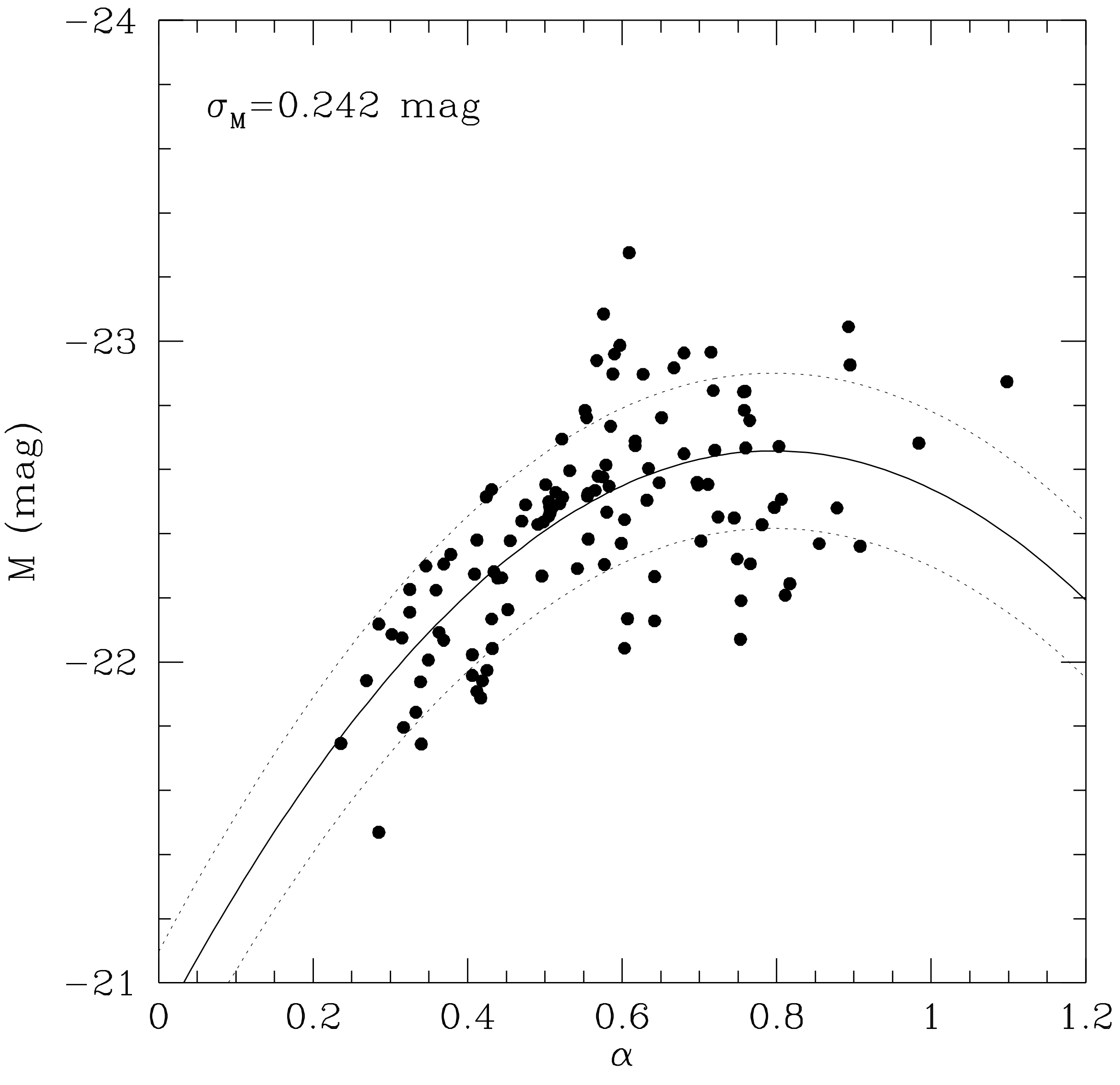

Fig. 5.— The $M$–$\alpha$ relation with the best-fitting quadratic shown as the solid line and the $\pm 1\sigma$ error range shown by the dotted lines.

confidence than LP and provide tighter constraints on cosmological models seeking to explain the amplitude of the observed bulk motion of the ACIF. In the next section we use Monte Carlo simulations to confirm these error estimates and also to check for biases in the various methods for deriving $\boldsymbol{v}_p$.

Our result for the motion of the ACIF w.r.t. the CMB, $\boldsymbol{F}{=}\boldsymbol{C}{-}\boldsymbol{v}_p$, is

$$\begin{aligned} V_x &= +513 \pm 188 \text{ km s}^{-1} \\ V_y &= -175 \pm 214 \text{ km s}^{-1} \\ V_z &= +631 \pm 164 \text{ km s}^{-1} \end{aligned} \tag{24}$$

which is of course also consistent with LP's result, though with 20% smaller uncertainties in each velocity component. This corresponds to $|\boldsymbol{F}|$=764 ± 160 in the direction $(l, b)$=(341,49) with a total angular error $\sigma_\theta = 20°$.

## 5. MONTE CARLO SIMULATIONS

We have constructed Monte Carlo simulations in order to test and compare the various methods for recovering the Local Group motion. The inputs to these simulations are the Local Group motion $\boldsymbol{v}_p$ and the $M$–$\alpha$ relation and its intrinsic scatter $\sigma_M$. From these and the observed directions, redshifts and $\alpha$'s of the BCGs we construct a simulated set of observed magnitudes. We then run our procedure to recover $\boldsymbol{v}_p$ from these simulated observations. The assumptions that we are making are: (i) since we use fixed $\alpha$'s, that these change little with different assumed values for $\boldsymbol{v}_p$ (as is in fact the case); (ii) that the distribution of the magnitude residuals really is Gaussian; and (iii) that the clusters themselves have negligible (or at least uncorrelated) peculiar motions.

For each method of recovering $\boldsymbol{v}_p$ we ran 5000 simulations using as inputs the best estimates of $\boldsymbol{v}_p$, $\sigma_M$ and the $M$–$\alpha$ relation given in the previous section. Figure 6 shows the distributions of the components of the recovered $\boldsymbol{v}_p$ and the distribution of the inferred $\sigma_M$ for each method. We find that when we maximize $\ln(\mathcal{L})$ or minimize $\sigma_M$ there is no significant bias in the recovered $\boldsymbol{v}_p$—the amplitude of the difference between the input $\boldsymbol{v}_p$ and the mean $\boldsymbol{v}_p$ recovered from the simulations is about 6 km s$^{-1}$, less than 1.5 times the joint standard error in the mean. When we minimize $v_c$, however, this difference is (+20,−33,−5), the amplitude of which is almost 9 times the joint standard error. The small but significant bias for this method results from the fact that one is mis-interpreting the magnitude residuals as peculiar velocities. Since the bias correction we derive for our preferred maximum likelihood method is both small and statistically insignificant, no such correction has been applied to the results given in the previous section.

It is interesting to note that the recovered value of the intrinsic scatter in the $M$-$\alpha$ relation, $\sigma_M$, is biased low by 0.006 mag. This reflects the fact that we are obliged to use the mean $M$-$\alpha$ relation calculated from the data rather than the unknown true relation, and hence tend to under-estimate the scatter. Thus our estimated scatter of 0.242 mag implies that the intrinsic scatter in the $M$-$\alpha$ relation is in fact 0.248 mag.

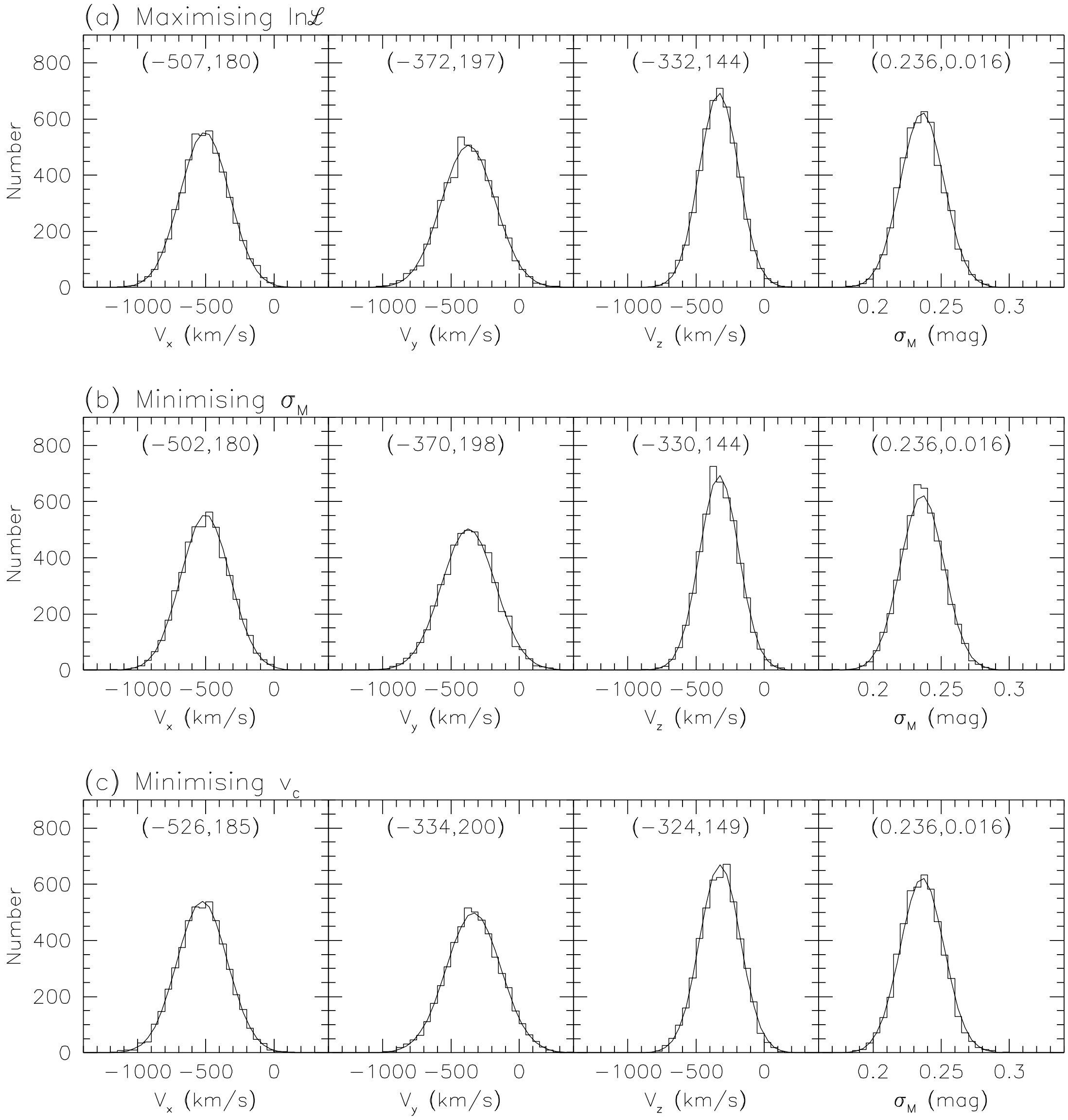

Fig. 6.— The components of the Local Group motion $\boldsymbol{v}_p$ and the scatter in the $M$–$\alpha$ relation $\sigma_M$ recovered from 5000 simulations by (a) maximising $\ln\mathcal{L}$, (b) minimising $\sigma_M$, and (c) minimising $v_c$. The input values for the simulations were $\boldsymbol{v}_p$=(−506,−367,−329) and $\sigma_M$=0.242 mag. The numbers in parentheses in each panel are the mean and standard deviation from the simulations; the smooth curve is the corresponding Gaussian.

As well as showing us that the maximum likelihood and minimum $\sigma_M$ methods are effectively unbiased, the simulations also demonstrate the reliability of the maximum likelihood estimates for the uncertainties. The standard deviations in the velocity components recovered from the simulations, (±180,±197,±144), are in good agreement with the uncertainties estimated by maximum likelihood, (±188,±213,±163). The slightly lower Monte Carlo errors are the consequence of the fact that in constructing the simulations we used the observed scatter of 0.242 mag rather than the true instrinsic scatter of 0.248 mag.

## 6. CONCLUSIONS

We have re-examined Lauer & Postman's (1994; LP) analysis of the Local Group motion w.r.t. the inertial frame defined by the mean motion of all the Abell clusters within 15,000 km s$^{-1}$ (the ACIF). Particular attention has been paid to the use of the $M$–$\alpha$ relation for brightest cluster galaxies (BCGs) as the distance estimator, and to the algorithm used to recover the peculiar motion of the Local Group w.r.t. the ACIF from these distances.

We derive an exact expression for the clusters' peculiar motions in terms of their magnitude residuals $\Delta M$ about the $M$–$\alpha$ relation. This result significantly improves on the approximation used by LP. We also show that, with the weighting scheme preferred by LP, obtaining an estimate of the Local Group motion by minimising the rms cluster peculiar velocity $v_c$ is closely related to a more straightforward minimisation of the rms residual about the $M$–$\alpha$ relation, $\sigma_M$. We argue that when the intrinsic scatter in the $M$–$\alpha$ relation dominates the scatter due to the clusters' peculiar velocities (as is the case here) then there is no physical basis for preferring to minimise $v_c$ rather than $\sigma_M$. In fact, mis-interpreting the magnitude residuals as due to peculiar velocities rather than intrinsic scatter results in a small bias in the derived Local Group motion. Using the observation that the magnitude residuals have a Gaussian distribution with intrinsic scatter $\sigma_M$=0.242 mag, we construct a maximum likelihood method for estimating the Local Group motion $\boldsymbol{v}_p$. This method has the advantage over the minimisation approaches that it provides direct estimates of the uncertainty in $\boldsymbol{v}_p$ and the confidence levels at which alternate values for $\boldsymbol{v}_p$ can be rejected, without recourse to Monte Carlo simulations.

We apply these methods to LP's original data to obtain an improved estimate of the Local Group motion. We find that the Local Group is moving w.r.t. the ACIF at $626 \pm 242$ km s$^{-1}$ towards $l$=216°, $b$=−28° (±20°). This implies that the ACIF is moving w.r.t. the CMB frame at $764 \pm 160$ km s$^{-1}$ towards $l$=341°, $b$=49° (±20°). These results are consistent with those of LP but have 20% smaller uncertainties. Calculations by Feldman & Watkins (1994) and Strauss et al. (1994) suggest that LP's original result for the bulk motion of the ACIF is inconsistent with most current cosmological models at a confidence level of 95% or higher. The slightly larger amplitude and significantly smaller uncertainties that we derive here for the motion of the ACIF w.r.t. the CMB further tighten the constraints on cosmological models.

A number of stimulating discussions with Alister Graham are gratefully acknowledged. Marc Postman provided a careful reading of the manuscript and his comments resulted in several material improvements.